\documentclass[11pt,a4paper]{article}

\usepackage[T1]{fontenc}
\usepackage[utf8]{inputenc}
\usepackage{mathptmx}
\usepackage[a4paper,left=24mm,right=24mm,top=22mm,bottom=24mm]{geometry}
\usepackage{microtype}
\usepackage{graphicx}
\usepackage{booktabs}
\usepackage{array}
\usepackage{caption}
\usepackage{subcaption}
\usepackage{xcolor}
\usepackage{hyperref}
\usepackage{fancyhdr}
\usepackage{enumitem}

\hypersetup{colorlinks=true,linkcolor=black,citecolor=black,urlcolor=blue}
\setlist{nosep}

\title{\bfseries Gamma irradiation of ATLAS18 ITk strip sensors\\affected by static charge}
\author{\parbox{0.96\textwidth}{\centering
M. Mikestikova\textsuperscript{$a,*$}, V. Fadeyev\textsuperscript{$b$},
P. Federicova\textsuperscript{$a$}, J. Fernandez-Tejero\textsuperscript{$c,d$},
C. Fleta\textsuperscript{$g$}, P. Gallus\textsuperscript{$e$},\\
C. Jessiman\textsuperscript{$f$}, J. Keller\textsuperscript{$f$},
C. Klein\textsuperscript{$f$}, T. Koffas\textsuperscript{$f$},
J. Kroll\textsuperscript{$a$}, J. Kvasnicka\textsuperscript{$a$},\\
E. Staats\textsuperscript{$f$}, P. Tuma\textsuperscript{$a$},
M. Ullan\textsuperscript{$g$}, and Y. Unno\textsuperscript{$h$}\\[0.6em]
\small\bfseries On behalf of the ATLAS ITk Collaboration\\[0.8em]
\raggedright\footnotesize
\textsuperscript{$a$}\textit{Institute of Physics, CAS, Na Slovance 2, 18200 Prague 8, Czech Republic}\\
\textsuperscript{$b$}\textit{Santa Cruz Institute for Particle Physics, University of California, Santa Cruz, CA 95064, USA}\\
\textsuperscript{$c$}\textit{Department of Physics, SFU, 8888 University Drive, Burnaby, B.C. V5A 1S6, Canada}\\
\textsuperscript{$d$}\textit{TRIUMF, 4004 Wesbrook Mall, Vancouver, B.C. V6T 2A3, Canada}\\
\textsuperscript{$e$}\textit{UJP PRAHA a.s., Nad Kam\'{i}nkou 1345, 15600 Prague--Zbraslav}\\
\textsuperscript{$f$}\textit{Physics Department, Carleton University, 1125 Colonel By Drive, Ottawa, Ontario, K1S 5B6, Canada}\\
\textsuperscript{$g$}\textit{IMB-CNM, CSIC, Campus UAB-Bellaterra, 08193 Barcelona, Spain}\\
\textsuperscript{$h$}\textit{Institute of Particle and Nuclear Study, KEK, 1-1 Oho, Tsukuba, Ibaraki 305-0801, Japan}\\[0.5em]
\textsuperscript{*}Corresponding author. E-mail: \href{mailto:marcela.mikestikova@cern.ch}{marcela.mikestikova@cern.ch}
}}
\date{}

\begin{document}
\maketitle
\thispagestyle{empty}
\vspace{-2.2em}

\begin{abstract}
Construction of the new all-silicon Inner Tracker (ITk), developed by the ATLAS collaboration to be able to track charged particles produced at the High-Luminosity LHC, started in 2021 and is expected to continue until 2028. The ITk detector will include approximately 18,000 highly segmented and radiation-hard $n^{+}$-in-$p$ silicon strip sensors, which are being manufactured by Hamamatsu Photonics. Upon their delivery, the ATLAS ITk strip sensor collaboration performs detailed measurements of sensors to monitor the quality of all fabricated pieces.

QC electrical tests include current--voltage (IV) and capacitance--voltage (CV) tests, full strip tests, and a measurement of the long-term stability of the sensor leakage current. While most sensors demonstrate excellent performance during QC testing, we have nevertheless observed that a number of sensors from several production batches failed the electrical tests.

Accumulated data indicate a strong correlation between observed electrical test failures and high electrostatic charge measured on the sensor surface during initial reception tests. This electrostatic charge enhances the risk of ``local trapped charge'' events during manufacturing, shipping, and handling procedures, resulting in failed electrical QC tests.

In the presented study, we have investigated whether the total ionizing dose (TID) expected in the real experiment can effectively resolve early breakdown or low interstrip isolation caused by the electrostatic charge. Selected charge-affected sensors were irradiated with gamma rays from a $^{60}$Co source for a number of TID values. The results of this study indicate that the negative effects of the electrostatic charge on the critical sensor characteristics disappear after a very small amount of accumulated TID, which corresponds to one or two days in the experiment. This finding gives us confidence in mitigating the issue of electrostatic charge during the operation of the ITk strip sensors in the real experiment.
\end{abstract}

\begin{flushleft}\small\itshape
The 32nd International Workshop on Vertex Detectors (VERTEX2023)\\
16--20 October 2023\\
Sestri Levante, Genova, Italy
\end{flushleft}

\section{Introduction}

The upgrade of the Large Hadron Collider (LHC) to the High-Luminosity LHC (HL-LHC), with peak instantaneous luminosity reaching $7.5\times10^{34}\,\mathrm{cm}^{-2}\mathrm{s}^{-1}$~\cite{itk-strip-tdr}, will require the replacement of the existing ATLAS Inner Detector with a new all-silicon tracker containing a new type and design of silicon sensors. The innermost part of the new Inner Tracker (ITk) will consist of pixel detectors, whereas the outer radii will be composed of strip detectors~\cite{itk-strip-tdr,itk-pixel-tdr}. Each individual module of the ITk strip detector includes a single-sided micro-strip sensor made with $n^{+}$-strips implanted on p-type silicon bulk ($n^{+}$-in-$p$). The ITk strip detector will consist of two types of barrel sensors with a square geometry of $9.8\times9.8\,\mathrm{cm}^{2}$, implementing short strips (24.1~mm) for the inner two barrel layers and long strips (48.3~mm) for the outer two barrel layers, and six types of endcap sensors with roughly trapezoidal shapes and strips forming a fan geometry. The endcap strip lengths vary from 15~mm to 60~mm depending on the radius. The pitch in a barrel sensor is 75.5~$\mu$m, while it varies around a mean value of 75~$\mu$m for the endcap sensors~\cite{unno2023}. Aluminum strips are AC-coupled to the implant strips, which are biased through polysilicon bias resistors. The ITk detector will include 17,888 strip sensors covering 165~m$^{2}$ with a total of approximately 60 million strips. Production of the total amount of 20,800 ITk strip sensors (including spares) started at HPK in 2021 and is scheduled for completion in 2025.

Upon delivery, the ATLAS ITk strip sensor collaboration performs detailed measurements of individual production sensor characteristics to monitor the quality of all fabricated devices. Several institutes involved in the complex testing program (Quality Control, or QC) conduct mechanical and electrical measurements of the sensors to ensure that their characteristics are within the specifications defined by the collaboration. QC electrical tests include current--voltage (IV) and capacitance--voltage (CV) tests conducted on each individual sensor. Additionally, a full strip test and a measurement of the long-term stability of the sensor leakage current are performed on a sample comprising 2--10\% of the total sensors. The full strip test checks the uniformity of electrical characteristics throughout the wafer surface. Each individual sensor strip is contacted and its impedance to ground is measured to identify the potential presence of metal shorts, broken implants, faulty bias resistors or low interstrip isolation, as well as pinholes or punch-throughs in the dielectrics. The current stability test is performed to check the sensor leakage current variation over tens of hours. During production, measurements of the surface electrostatic charge were added for each individual sensor.

By now, approximately 50\% of production sensors have been delivered and tested. While most sensors have demonstrated high-quality performance in QC tests, several sensors from different production batches failed the electrical QC tests. Early breakdowns were observed in IV tests, where the breakdown voltage ($V_{\mathrm{BD}}$) was measured to be below the required 500~V, whereas areas with low interstrip isolation in full strip tests as well as current instabilities during the long-term current stability studies were also observed. Additionally, on some occasions, IV failures were observed after the long-term leakage current stability test.

Accumulated data indicate a strong correlation between observed electrical failures and a high electrostatic charge, resulting in a potential of several hundreds of volts, measured on the surface of the sensors and of the plastic sheets used for sensor mechanical protection during the initial reception test. The electrostatic charge enhances the occurrence of ``local trapped charge'' events during handling procedures, manufacturing steps, and shipping, resulting in failed electrical QC tests. Such an event is caused, for example, by touching a charged sensor surface with the vacuum stencil or another tool. In such cases, the charge becomes trapped at the point of contact. Electrostatic charge accumulated on the surface of the sensor in the strip area can lead to the loss of interstrip isolation, whereas charge accumulated close to the edge and bias ring can cause instability of the leakage current or low breakdown voltage. To mitigate the above-described issues, the QC testing institutes modified the sensor handling procedures and introduced sensor recovery techniques, such as irradiating the sensor surface with UV-A (315--400~nm wavelength) or UV-C light (100--280~nm), applying an intensive flow of ionized gas, or subjecting the sensors to high-temperature exposure at 150~$^{\circ}$C for 16~h. In many cases, simply storing the sensors in a dry environment for a few days or weeks helps to cure the sensors. Additional information about sensor recovery techniques can be found in Ref.~\cite{federicova2023}. However, despite the implementation of these methods, there is still a possibility that some affected sensors may go unnoticed during the QC testing process, or that ``local trapped charge'' events could occur during later manipulation of the sensor.

The question therefore arises: will the sensor issues caused by electrostatic charge be ``cured'' by ionization in the real experiment, and if so, how quickly? To investigate whether the total ionizing dose (TID) expected in the real experiment can effectively resolve early breakdown or low interstrip isolation caused by the electrostatic charge, the affected sensors were irradiated with gamma rays from a $^{60}$Co source. To determine the TID and the corresponding duration of the real experiment required for such a cure, irradiation with several TID values was planned.

\section{Sensor selection and irradiation}

In this experiment, we chose three sensors that exhibited early breakdown in IV tests and three sensors with a low interstrip isolation area identified in the full strip test for the gamma irradiation study. Our plan was to irradiate only two from each category (early breakdown and low interstrip isolation) and keep one sensor from each category as a reference sample. The reference sample would undergo no irradiation but would be subjected to identical conditions in terms of transport, storage, and handling as the irradiated sensors. The selected sensors, along with their breakdown voltage ($V_{\mathrm{BD}}$) and the number of strips with low interstrip isolation, are given in Table~\ref{tab:samples}. Initial IV tests were conducted as part of the standard sensor QC testing and were repeated after two months of storage in a drying cabinet, just before irradiation.

\begin{table}[htbp]
\centering
\caption{Selected samples for the irradiation study. R1 and R3 indicate the sensor type and Wxxx represents the wafer number. Samples highlighted in red with an ``R'' denote the reference sample.}
\label{tab:samples}
\small
\begin{tabular}{@{}l>{\centering\arraybackslash}p{3.3cm}l>{\centering\arraybackslash}p{4.2cm}@{}}
\toprule
Sample & Initial $V_{\mathrm{BD}}$ / $V_{\mathrm{BD}}$ just before irradiation [V] & Sample & Low interstrip isolation area: affected strips in initial test / just before irradiation \\
\midrule
R1-W607 & 150 / 160 & R1-W635 & 40 / 40 \\
R1-W617 & 150 / 150 & \textcolor{red}{R1-W620-R} & \textcolor{red}{26 / 18} \\
\textcolor{red}{R1-W650-R} & \textcolor{red}{150 / 200} & R3-W1014 & 28 / 24 \\
\bottomrule
\end{tabular}
\end{table}

\begin{figure}[htbp]
\centering
\begin{subfigure}[t]{0.48\textwidth}\centering
\includegraphics[width=\linewidth]{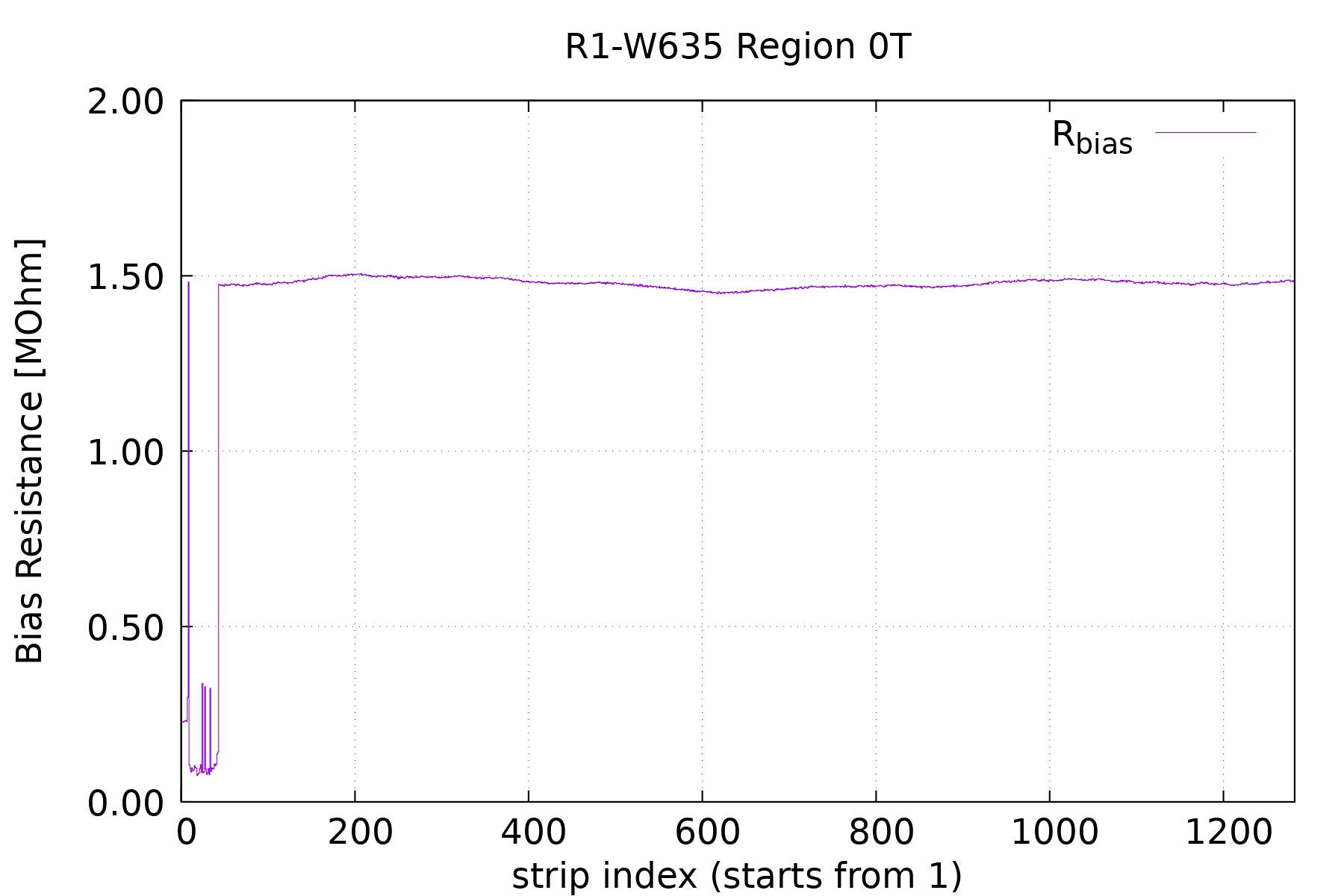}
\end{subfigure}\hfill
\begin{subfigure}[t]{0.48\textwidth}\centering
\includegraphics[width=\linewidth]{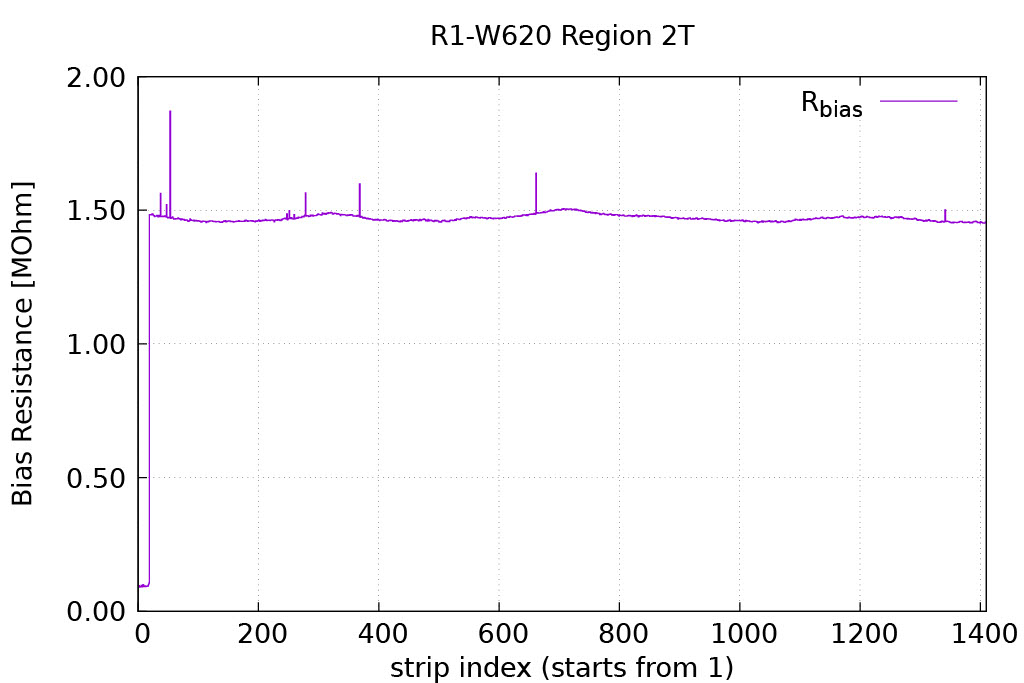}
\end{subfigure}
\caption{Full strip tests of R1-W635 (left) and R1-W620-R (right) sensors measured after two months of dry storage, i.e. just before irradiation.}
\label{fig:before}
\end{figure}

Figure~\ref{fig:before} shows the full strip test results of bias resistance ($R_{\mathrm{BIAS}}$) of the R1-W635 and R1-W620-R sensors tested just before planned irradiation. The low interstrip isolation manifests as areas with low $R_{\mathrm{BIAS}}$ values in the full strip test measuring setup. The required value of $R_{\mathrm{BIAS}}$ according to the ATLAS specification is $1.5\pm0.5\,\mathrm{M}\Omega$.

The target TID values chosen for this experiment were 11, 49, 195, 590, and 1200~krad, corresponding approximately to one week, one month, six months, one year, and two years of sensor exposure in the real experiment according to the simulations~\cite{radiation-simulation}. The worst-case scenario for curing sensors affected by electrostatic charge is in regions in the ITk detector where the TID is minimal, as it would take longer to get sensors cured in these areas. Specifically, this applies to the central region of the outermost barrel cylinder, with a 5.9~Mrad target dose including a safety factor of 1.5.

The sensors were irradiated by a $^{60}$Co gamma radionuclide source installed at UJP PRAHA a.s.~\cite{ujp} with a dose rate of 1~krad/min. The sensors were irradiated individually, one by one, in a custom-made charged-particle equilibrium (CPE) box constructed with an outer layer of 1.5~mm of Pb and an inner layer of 1~mm of Al, following the recommendations of ESCC~\cite{escc}. While using the CPE box during the irradiation, dose enhancement from low-energy scattered radiation is minimized by producing electron equilibrium, and a uniform distribution of energy deposited in the sample is ensured. The reference samples were physically moved inside the CPE box and taken out, but were not subjected to irradiation.

\section{Results}

\begin{figure}[htbp]
\centering
\begin{subfigure}[t]{0.48\textwidth}\centering
\includegraphics[width=\linewidth]{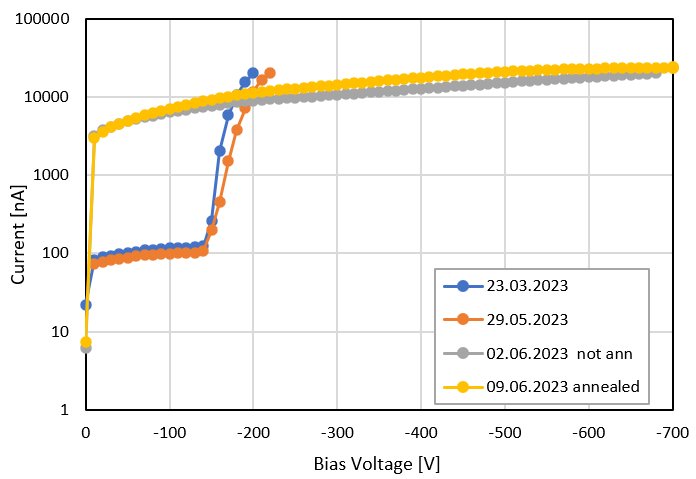}\caption{}\label{fig:iv-a}
\end{subfigure}\hfill
\begin{subfigure}[t]{0.48\textwidth}\centering
\includegraphics[width=\linewidth]{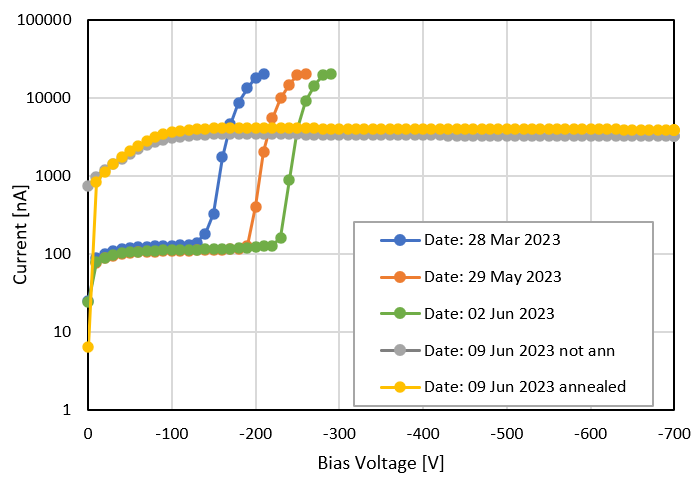}\caption{}\label{fig:iv-b}
\end{subfigure}
\caption{IV characteristics measured before and after gamma irradiation for R1-W617 (a) and for the reference sample R1-W650-R (b).}
\label{fig:iv}
\end{figure}

The measurements of IV characteristics after irradiation with a TID of 11~krad for sample R1-W617 are shown in Figure~\ref{fig:iv}a. The IV characteristics measured before irradiation, with a breakdown voltage of 150~V, are also displayed. Early breakdown (BD) of sensors W617 and W607 (not shown in the plot) was no longer present after gamma irradiation to 11~krad. In Figure~\ref{fig:iv}b, IV characteristics of the reference sample R1-W650-R are shown. The reference sensor R1-W650-R, which was transported and handled alongside the other sensors and was not subjected to irradiation, did not recover, as seen from the green line corresponding to the measurement on June 2, 2023. Its breakdown voltage increased only slightly, to 250~V, in comparison to the initial measurements. The leakage current of sensors irradiated to 11~krad (R1-W607 and R1-W617) increased by two orders of magnitude. Such an increase in current is expected since the measurement was performed at room temperature. In the real experiment, sensors will be operated at temperatures below $-30\,^{\circ}$C. According to the previous study~\cite{zatocilova2023}, the increase of the total current, which comprises both surface and bulk components, after gamma irradiation is due to the increase of the surface part of the total current. The bulk leakage current increases only very little after a TID as small as 11~krad. Annealing at 60~$^{\circ}$C for 80 minutes did not decrease the current.

As gamma irradiation resolved issues successfully during the initial planned step of irradiation with a TID of 11~krad (approximately one week in the real experiment), no further irradiations to higher TID values were needed. Instead, the reference samples were irradiated to an even lower TID of 1.5~krad, corresponding to one day in the real experiment. Figure~\ref{fig:iv}b shows measurements performed after gamma irradiation of the reference sample R1-W650-R to 1.5~krad (gray and yellow lines measured on June 9). The early BD observed before irradiation was no longer present even after this small irradiation. Annealing again did not help to decrease the leakage current in this case.

\begin{figure}[htbp]
\centering
\begin{subfigure}[t]{0.48\textwidth}\centering
\includegraphics[width=\linewidth]{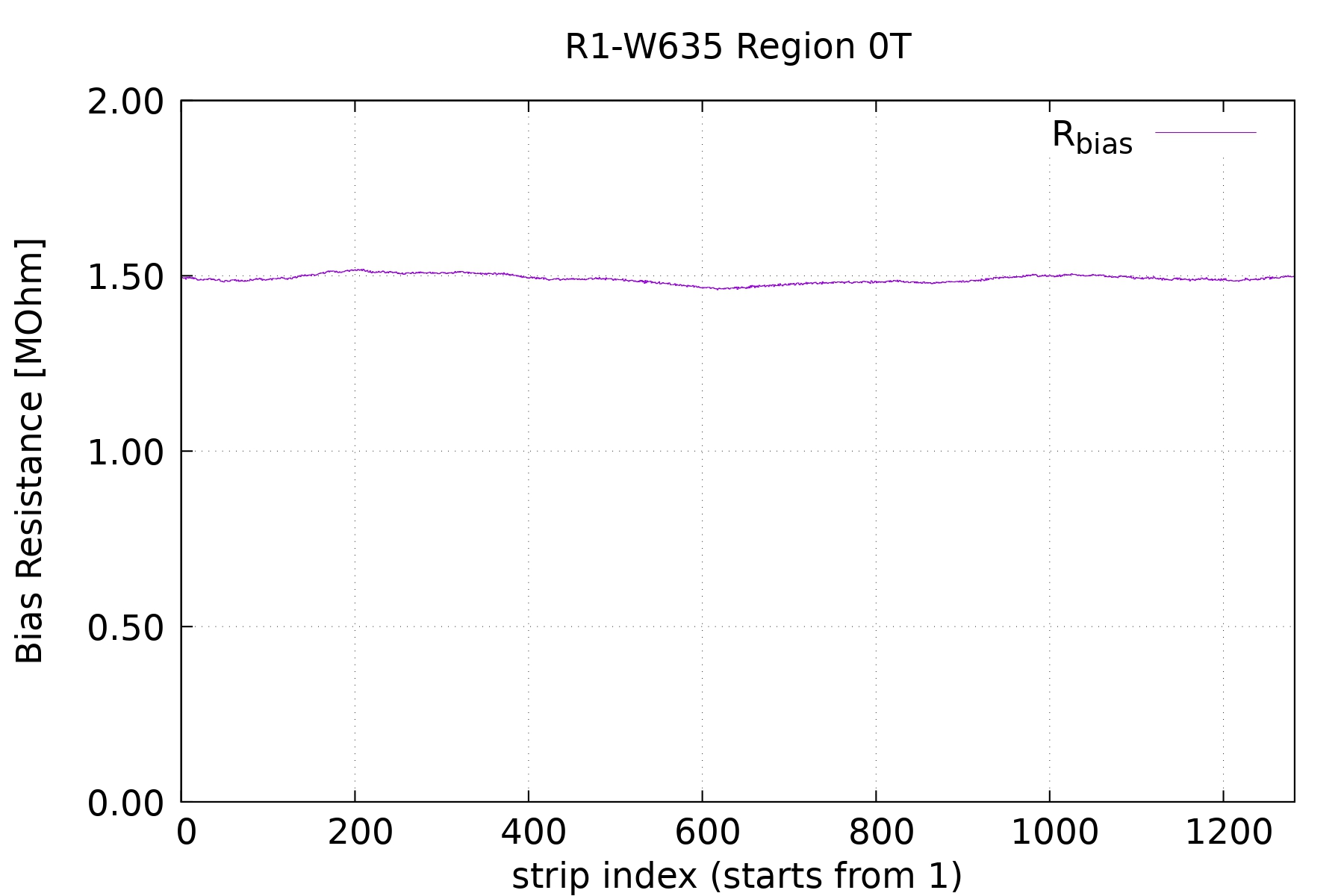}
\end{subfigure}\hfill
\begin{subfigure}[t]{0.48\textwidth}\centering
\includegraphics[width=\linewidth]{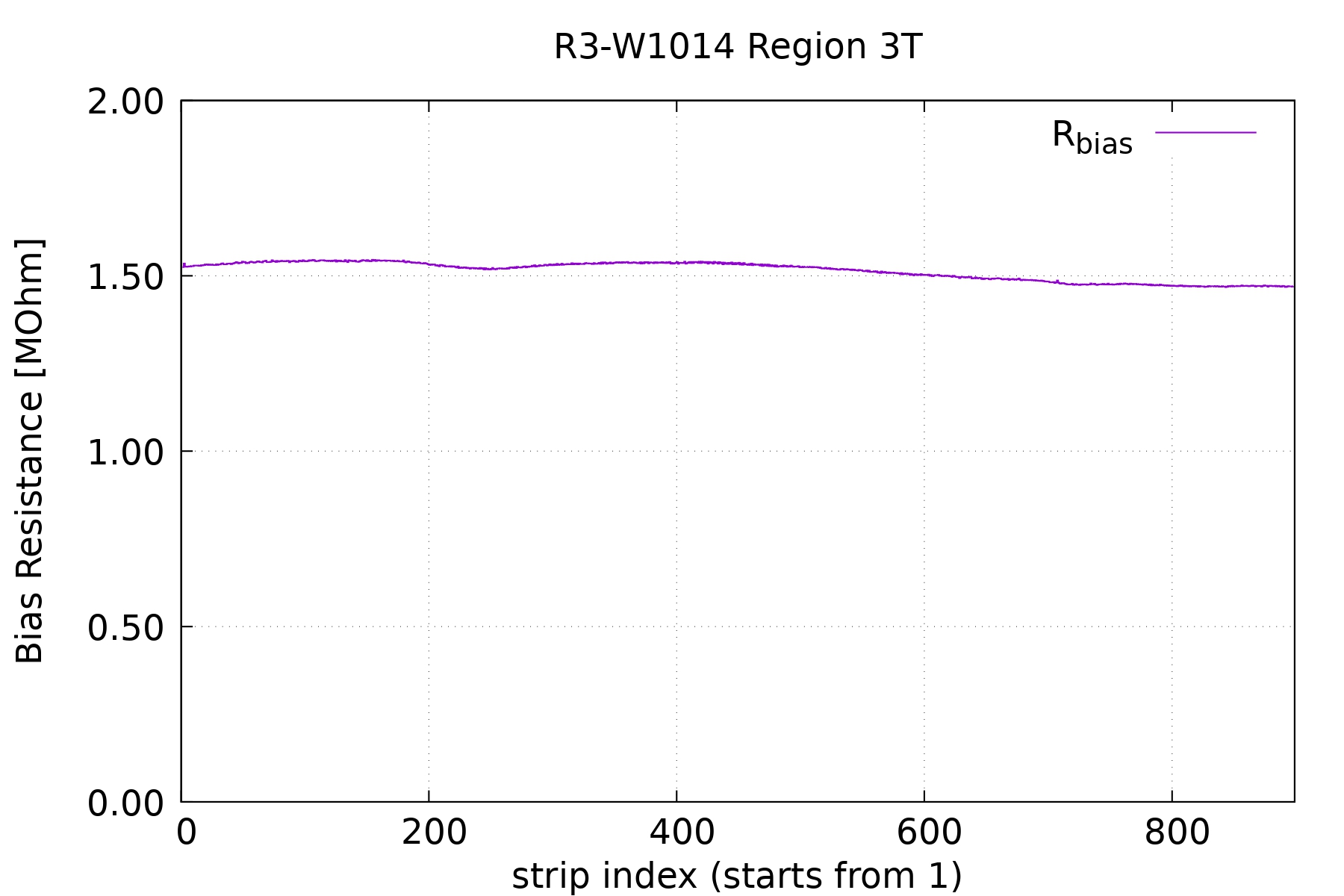}
\end{subfigure}
\caption{Full strip tests of R1-W635 (left) and R3-W1014 (right) measured after gamma irradiation of 11~krad.}
\label{fig:after11}
\end{figure}

\begin{figure}[htbp]
\centering
\begin{subfigure}[t]{0.48\textwidth}\centering
\includegraphics[width=\linewidth]{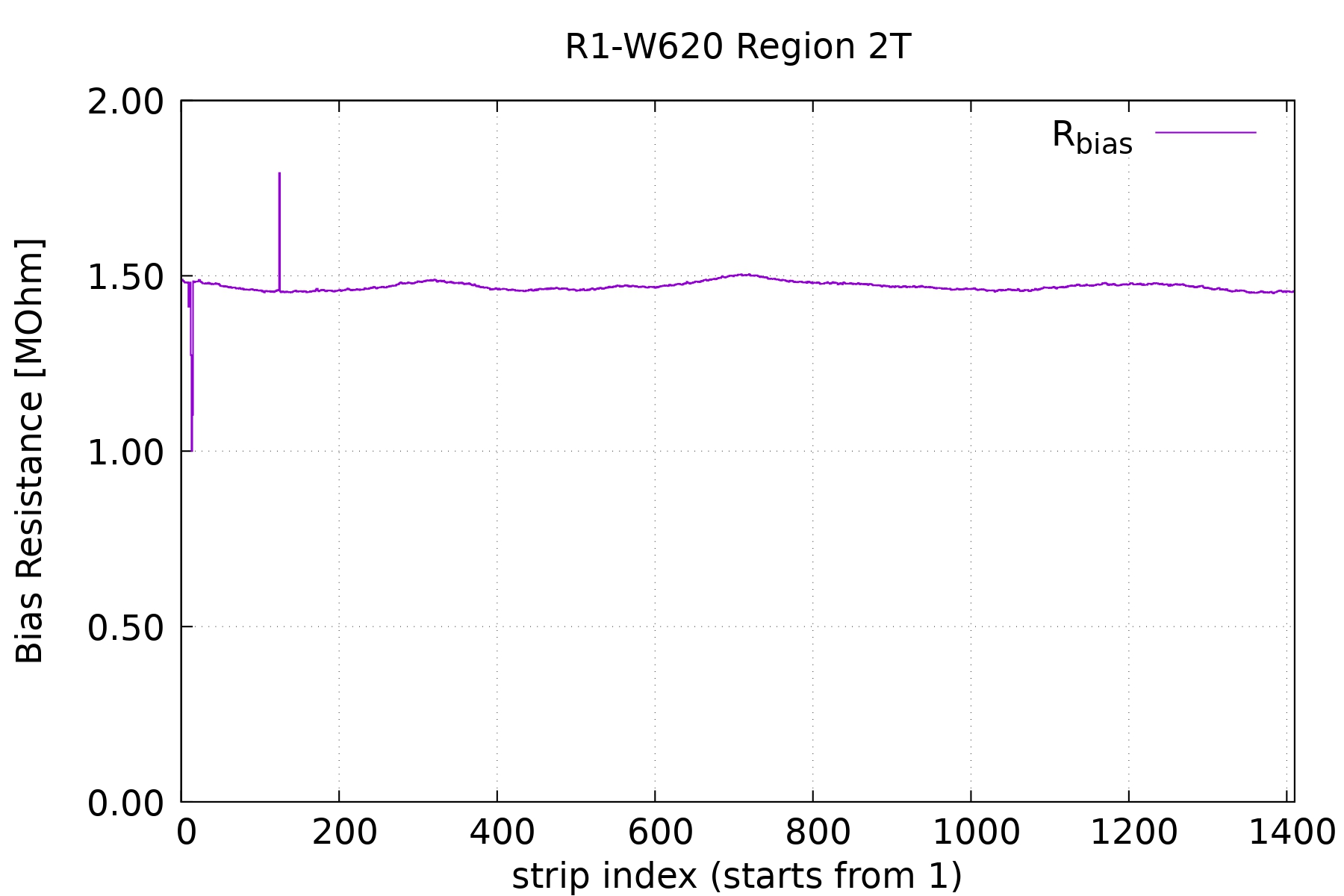}
\end{subfigure}\hfill
\begin{subfigure}[t]{0.48\textwidth}\centering
\includegraphics[width=\linewidth]{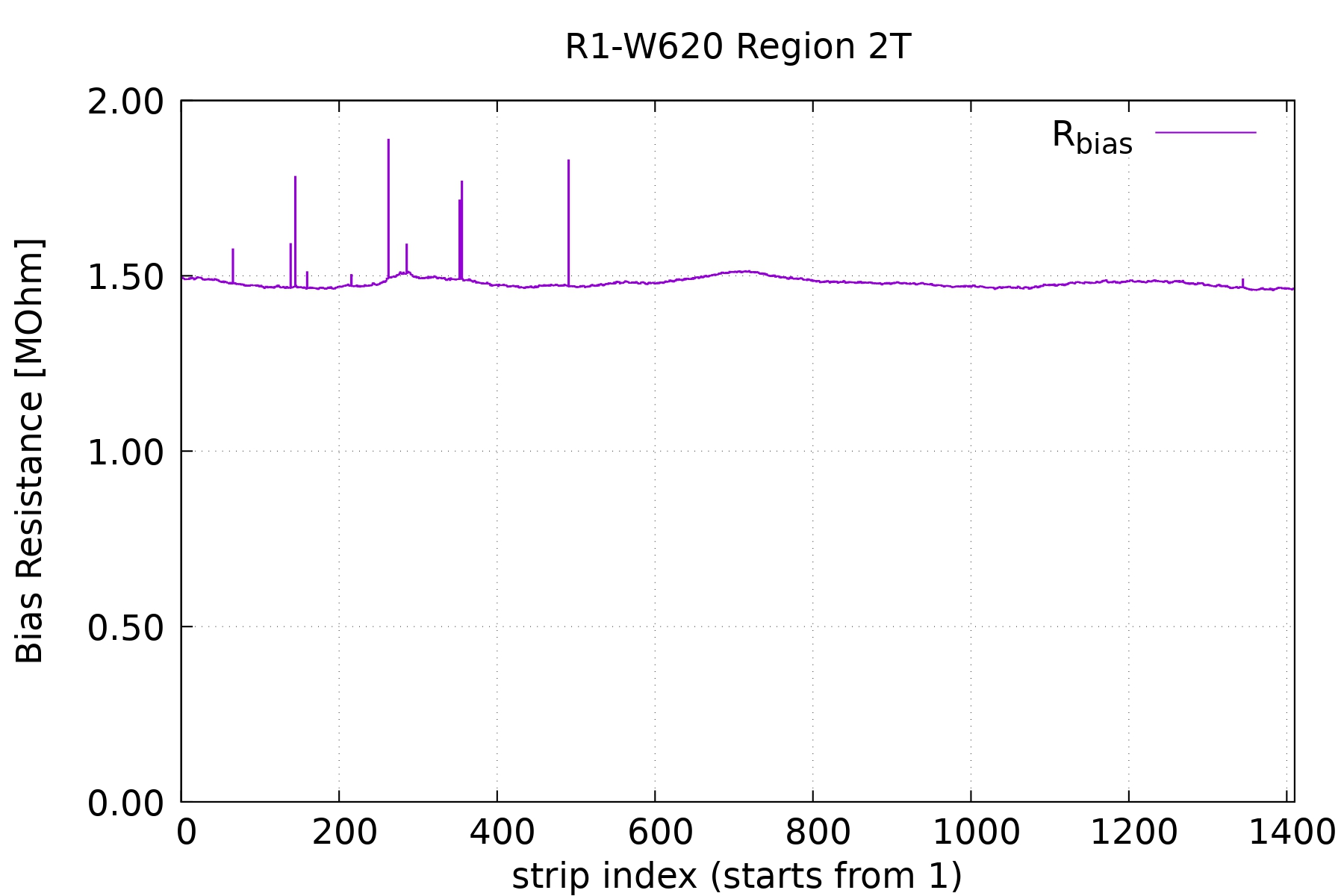}
\end{subfigure}
\caption{Full strip tests of the R1-W620-R sensor measured after a total delivered ionizing dose of 1.5~krad (left) and 3~krad (right).}
\label{fig:afterlow}
\end{figure}

The full strip tests done after irradiation to a TID of 11~krad are shown in Figure~\ref{fig:after11}. The gamma irradiation with a TID of 11~krad completely cured the low interstrip isolation areas on sensors R1-W635 and R3-W1014. Figure~\ref{fig:afterlow} shows the reference sample R1-W620-R after TIDs of 1.5 and 3~krad. The reference sample R1-W620-R irradiated only to 1.5~krad has a bias resistance value of the affected strips just above the required limit of 1~M$\Omega$, but it was not fully recovered. An additional gamma irradiation with a TID of 1.5~krad was thus applied, and the results are shown in Figure~\ref{fig:afterlow} (right). After the TID of 3~krad, the area of low interstrip isolation, manifested by a lower bias resistance value, disappeared. The few strips with increased bias resistance are issues related to non-ideal probe contact.

\section{Conclusions}

The results of this study indicate that the negative effects of the electrostatic charge accumulated on the sensor surface completely disappear after a very small amount of delivered total ionizing dose, which corresponds to one or two days of operation in the real experiment. These findings give us confidence that any issues related to the accumulation of electrostatic charge on the ITk sensor surface will be very quickly mitigated under the conditions of the real experiment.

\section*{Acknowledgements}

This work was supported by the Ministry of Education, Youth and Sports of the Czech Republic through projects LM2023040 CERN-CZ and LTT17018 Inter-Excellence.


\begin{thebibliography}{99}
\bibitem{itk-strip-tdr} ATLAS Collaboration, \emph{Technical Design Report for the ATLAS Inner Tracker Strip Detector}, CERN-LHCC-2017-005; ATLAS-TDR-025.
\bibitem{itk-pixel-tdr} ATLAS Collaboration, \emph{Technical Design Report for the ATLAS Inner Tracker Pixel Detector}, CERN-LHCC-2017-021; ATLAS-TDR-030.
\bibitem{unno2023} Y. Unno et al., \emph{Specifications and Pre-Production of $n^{+}$-in-$p$ Large-format Strip Sensors fabricated in 6-inch Silicon Wafers, ATLAS18, for Inner Tracker of ATLAS Detector for High-Luminosity Large Hadron Collider}, JINST \textbf{18} T03008 (2023).
\bibitem{federicova2023} P. Federicova et al., \emph{Setups for eliminating static charge of the ATLAS18 strip sensors}, submitted to JINST as iWoRiD2023 proceedings.
\bibitem{radiation-simulation} ATLAS Collaboration, \emph{Radiation Simulation}, \url{https://twiki.cern.ch/twiki/bin/view/AtlasPublic/RadiationSimulationPublicResults#Phase_II_Upgrade_Mar_2018_AN1}.
\bibitem{ujp} UJP PRAHA a.s., \url{https://ujp.cz/en/}.
\bibitem{escc} European Space Agency, \emph{Total Dose Steady-State Irradiation Test Method}, ESCC Basic Specification No. 22900 (2016).
\bibitem{zatocilova2023} I. Zatocilova et al., \emph{Study of bulk damage of high dose gamma irradiated p-type silicon diodes with different resistivities}, submitted to JINST as iWoRiD2023 proceedings.
\end{thebibliography}
\end{document}